\documentclass[a4paper,11pt]{article}
\usepackage{pos}

\usepackage{subfigure}

\title{The flux tube profile in full QCD}

\author[a]{Marshall Baker}
\author[b]{Volodymyr Chelnokov}
\author*[c]{Leonardo Cosmai}
\author[b]{Francesca Cuteri}
\author[d,e]{Alessandro Papa}

\affiliation[a]{Department of Physics, University of Washington, 3910 15th Ave. NE, Seattle, WA 98195-1560, USA}

\affiliation[b]{Institut f\"ur Theoretische Physik, Goethe Universit\"at, Max-von-Laue-Str. 160438 Frankfurt am Main, Germany}

\affiliation[c]{INFN - Sezione di Bari, via Amendola 173,  I-70126 Bari, Italy}

\affiliation[d]{INFN - Gruppo collegato di Cosenza, I-87036 Arcavacata di Rende, Cosenza, Italy}

\affiliation[e]{Dipartimento di Fisica dell'Universit\`a della Calabria, via Pietro Bucci, 87036 Arcavacata di Rende (CS), Italy}

\emailAdd{leonardo.cosmai@ba.infn.it}
\emailAdd{ mbaker4@uw.edu}
\emailAdd{chelnokov@itp.uni-frankfurt.de}
\emailAdd{cuteri@itp.uni-frankfurt.de}
\emailAdd{alessandro.papa@fis.unical.it}

\abstract{
We measure the spatial distribution of all components of the color fields surrounding a static quark antiquark pair in QCD with (2+1) HISQ flavors.
We isolate the nonperturbative component of the longitudinal chromoelectric color field responsible for the linear term in the confining potential.
}

\FullConference{%
 The 38th International Symposium on Lattice Field Theory, LATTICE2021
  26th-30th July, 2021
  Zoom/Gather@Massachusetts Institute of Technology
}

\begin{document}
\maketitle

\section{Introduction}
Even though Quantum chromodynamics (QCD) is universally accepted as the theory of strong interactions,
color confinement still remains an unsolved problem. Although the well established phenomenon of the confinement of quarks and gluons inside hadrons is without any doubt encoded into the QCD Lagrangian, our
current understanding relies only on a number of models of the QCD vacuum (for a review, see Refs.~\cite{Greensite:2011zz,Diakonov:2009jq} ). A theoretical a priori explanation of the area law for large Wilson loops, resulting from a long distance linear confining quark-antiquark potential, is still missing.

First-principle numerical Monte Carlo simulations of QCD on a space-time lattice represent the most eligible tool for testing (or ruling out) models of confinement, but can also provide ``phenomenological'' results that can suggest new insights into
the mechanism of confinement. Numerical simulations have clearly established the presence of a linear confining potential between a static quark an antiquark for distances larger than 0.5 fm up to infinite distance in SU(3) pure gauge theory, and  up to a distance of about 1.4 fm in presence of dynamical quarks,
where {\em string breaking} should take place~\cite{Philipsen:1998de,Kratochvila:2002vm,Bali:2005fu,Koch:2015qxr,Koch:2018puh,KnechtlyLat21}.
This long-distance linear potential is naturally associated with the observation of a tube-like structure of the chromoelectric field 
produced by a static quark-antiquark pair. 

A large amount of numerical investigations of flux tubes has accumulated in SU(2) and SU(3) Yang-Mills theories~\cite{Fukugita:1983du,Kiskis:1984ru,Flower:1985gs,Wosiek:1987kx,DiGiacomo:1989yp,DiGiacomo:1990hc,Cea:1992sd,Matsubara:1993nq,Cea:1994ed,Cea:1995zt,Bali:1994de,Green:1996be,Skala:1996ar,Haymaker:2005py,D'Alessandro:2006ug,Cardaci:2010tb,Cea:2012qw,Cea:2013oba,Cea:2014uja,Cea:2014hma,Cardoso:2013lla,Caselle:2014eka,Cea:2015wjd,Cea:2017ocq,Shuryak:2018ytg,Bonati:2018uwh,Shibata:2019bke} mostly
aimed at studying the shape of the chromoelectric field on the transverse plane at the midpoint of the line connecting the static quark and antiquark, given that the other two components of the chromoelectric field and all the three components of the chromomagnetic field are suppressed in that plane. The last few years witnessed several  numerical efforts toward a more thorough description of the color fields around static sources.  In this regard measurements of all components of both chromoelectric an chromomagnetic fields on all transverse planes passing through the line between quarks  were performed~\cite{Baker:2018mhw,Baker:2019gsi};  other contributions come from the study of the spatial distribution of the stress energy momentum tensor~\cite{Yanagihara:2018qqg,Yanagihara:2019foh} and of the flux densities for hybrid static potentials~\cite{Bikudo:2018,Mueller:2019mkh}. 

In our numerical studies~\cite{Baker:2018mhw,Baker:2019gsi} of the spatial distribution in three dimensions of all components of the chromoelectric and chromomagnetic fields generated by 
a static quark-antiquark pair in pure SU(3) lattice gauge theory, we found that the larger contribution is given by longitudinal chromoelectric field, but still the transverse chromoelectric components are large enough to be fit to a Coulomb-like ``perturbative''  field, whereas all the chromomagnetic components of the color field are compatible with zero within the statistical uncertainties.
We also introduced a new procedure to extract this perturbative Coulomb field from the transverse components of the chromoelectric field that
allows to isolate the nonperturbative confining field. From the knowledge of the nonpreturbative part of the longitudinal chromoelectric field we were able to extract some relevant 
parameters of the confining flux tube, such as the mean width and the string tension. 
In this paper we present preliminary results obtained in the case of QCD with (2+1) HISQ~\cite{Follana:2006rc,MILC:2009mpl} flavors.

\section{Lattice setup and numerical results}

The lattice operator whose vacuum expectation value gives us access to the components of the color field generated by a static $q \bar q$ pair is the following connected correlator~\cite{DiGiacomo:1989yp,DiGiacomo:1990hc,Kuzmenko:2000bq,DiGiacomo:2000va}:
\begin{equation}
    \label{rhoW}
    \rho_{W,\,\mu\nu}^{\rm conn} = \frac{\left\langle {\rm tr}
      \left( W L U_P L^{\dagger} \right)  \right\rangle}
        { \left\langle {\rm tr} (W) \right\rangle }
        - \frac{1}{N} \,
        \frac{\left\langle {\rm tr} (U_P) {\rm tr} (W)  \right\rangle}
             { \left\langle {\rm tr} (W) \right\rangle } \; .
\end{equation}

\begin{figure}[tb] 
\centering
\subfigure[]{\includegraphics[width=0.4\textwidth]{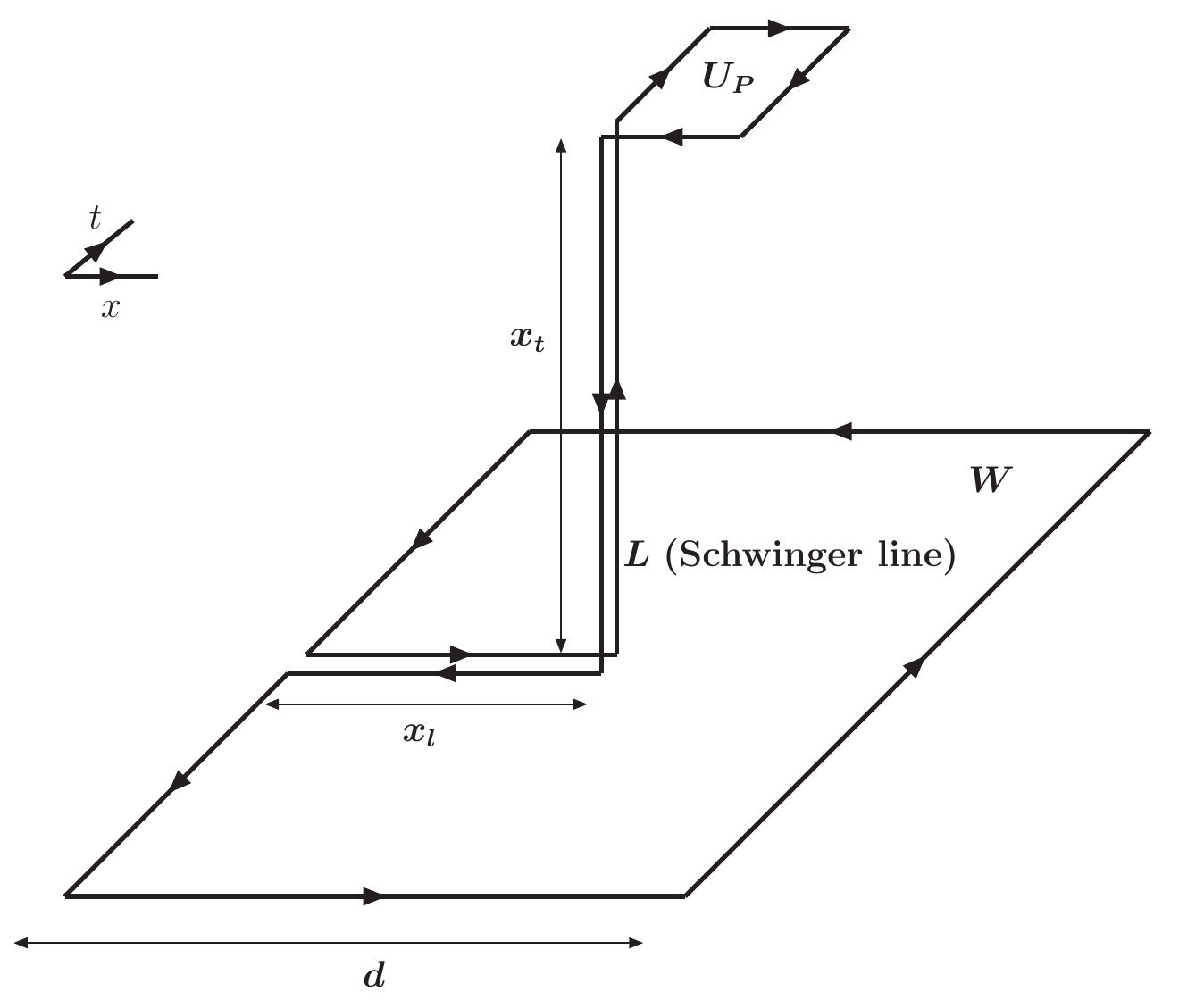}} 
\subfigure[]{\includegraphics[width=0.2\textwidth]{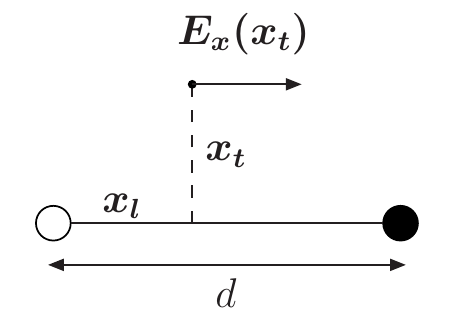}} 
\caption{
(a) The connected correlator given in Eq.~(\protect\ref{rhoW}) between the plaquette $U_{P}$ and the Wilson loop
(subtraction in $\rho_{W,\,\mu\nu}^{\rm conn}$ not explicitly drawn). (b) The longitudinal 
chromoelectric field $E_x(x_t)$ relative to the position of the static sources (represented by the white and black circles), 
for a given value of the transverse distance $x_t$ and of the longitudinal distance $x_l$.}
\label{fig:op_W}
\end{figure}

Here $U_P=U_{\mu\nu}(x)$ is the plaquette in the $(\mu,\nu)$ plane, connected to the Wilson loop $W$ by a Schwinger line $L$, and $N$ is the number of colors (see Fig.~\ref{fig:op_W}).
The correlation function defined in Eq.~(\ref{rhoW}) measures the field strength $F_{\mu\nu}$, since in the naive continuum
limit~\cite{DiGiacomo:1990hc}
\begin{equation}
    \label{rhoWlimcont}
    \rho_{W,\,\mu\nu}^{\rm conn}\stackrel{a \rightarrow 0}{\longrightarrow} a^2 g 
    \left[ \left\langle
      F_{\mu\nu}\right\rangle_{q\bar{q}} - \left\langle F_{\mu\nu}
      \right\rangle_0 \right]  \;,
\end{equation}
where $\langle\quad\rangle_{q \bar q}$ denotes the average in the presence of a static $q \bar q$ pair, and $\langle\quad\rangle_0$ is the vacuum average.
This relation is a necessary consequence of the gauge-invariance of the operator defined in Eq.~(\ref{rhoW}) and of its linear dependence on the color field in the continuum limit (see Ref.~\cite{Cea:2015wjd}).
The lattice definition of the quark-antiquark field-strength tensor
$F_{\mu\nu}$ is then obtained by equating the two sides of
 Eq.~(\ref{rhoWlimcont}) for finite lattice spacing (more details in Refs.~\cite{Baker:2018mhw,Baker:2019gsi}).

We carried out numerical simulations for QCD with (2+1) HISQ flavors on lattices $24^4$, $32^4$, and $48^4$ and for 
gauge coupling constants $\beta=6.445$, $\beta=7.158$, and $\beta=6.885$, respectively.  For each setup, we measured the color fields corresponding to two different quark-antiquark separations. The physical scale is the ``$r_1$-scale'' as defined in Ref.~\cite{Bazavov:2011nk}. We take the quark mass parameters in order to be on the line of constant physics determined by fixing the strange quark mass to its physical value $m_s$ at each value of the gauge coupling $\beta$. The light-quark mass has
been fixed at $m_l=m_s/20$. This corresponds to a pion mass $m_\pi=160$~MeV~\cite{Bazavov:2011nk}.

 The operator in Eq.~(\ref{rhoW}) undergoes a non-trivial renormalization~\cite{Battelli:2019lkz}, which depends on $x_t$. As discussed in Refs.~\cite{Baker:2018mhw,Baker:2019gsi}, 
 comparing our results with those in Ref.~\cite{Battelli:2019lkz} we concluded that smearing behaves as an effective renormalization, even though no analysis of the systematics of our approach as compared to theirs was ever done.
 
\begin{figure}[tb] 
\centering
\includegraphics[width=0.7\textwidth]{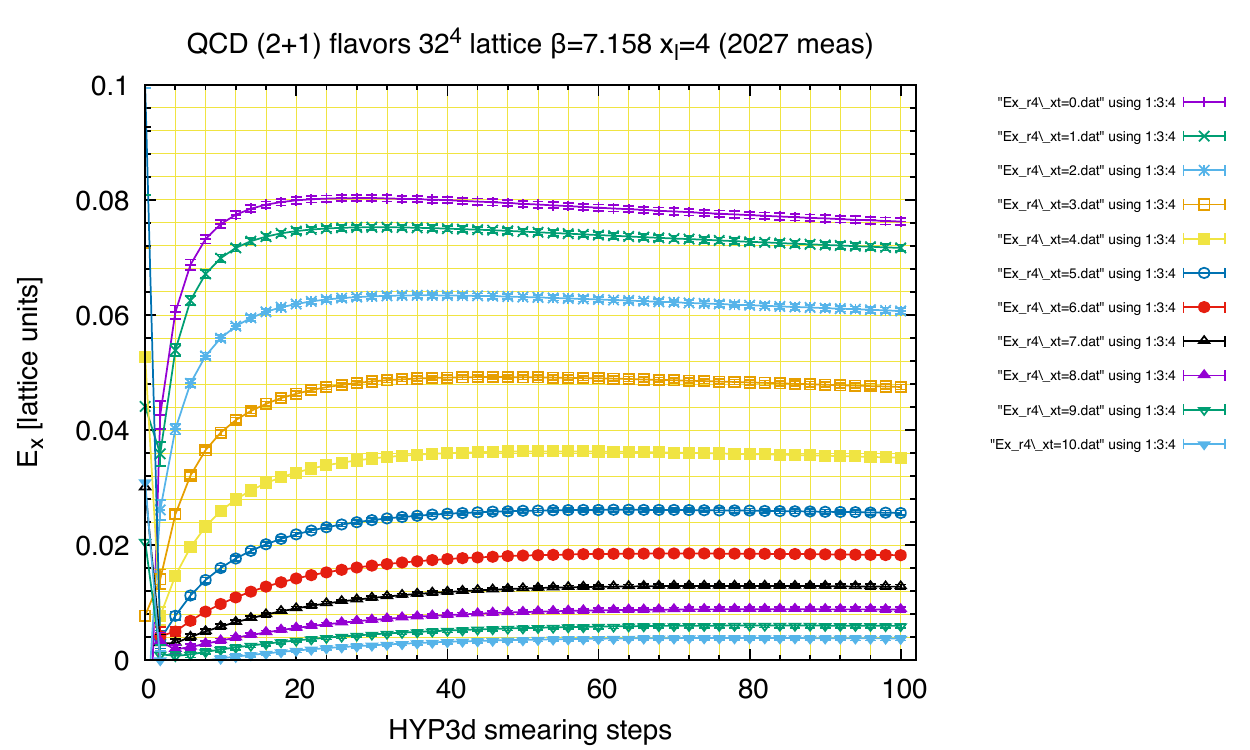} 
\caption{The $E_x(x_l,x_t)$ field in lattice units measured after 1 HYPt smearing versus the number of HYP3d smearings.
The values of $E_x(x_l,x_t)$ are displayed at a fixed value $x_l=4a$ for each value of the transverse distance $0\le x_t \le10$.}
\label{fig:smearing}
\end{figure}

 The smearing procedure can be validated {\it a posteriori} by the observation of continuum scaling, that is by checking that fields obtained in the same {\em physical} setup, but at different values of the coupling and of the optimal number of smearing step, are in good agreement in the range of gauge couplings used.
In the present work we smoothed gauge configurations using $1$ step of HYP~\cite{Hasenfratz:2001hp} smearing on temporal links
In Fig.~\ref{fig:smearing} we show an example of the smearing procedure applied in the case of measurements of the longitudinal component of the chromoelectic field on a $32^4$ lattice, $\beta=7.158$, and a distance of 0.75~fm between static sources at $x_l=4$, and with as many datasets as there are $x_t$ at which we measured the field'.

\begin{figure}[tb] 
\centering
\subfigure[]{\includegraphics[width=0.4\textwidth,clip]{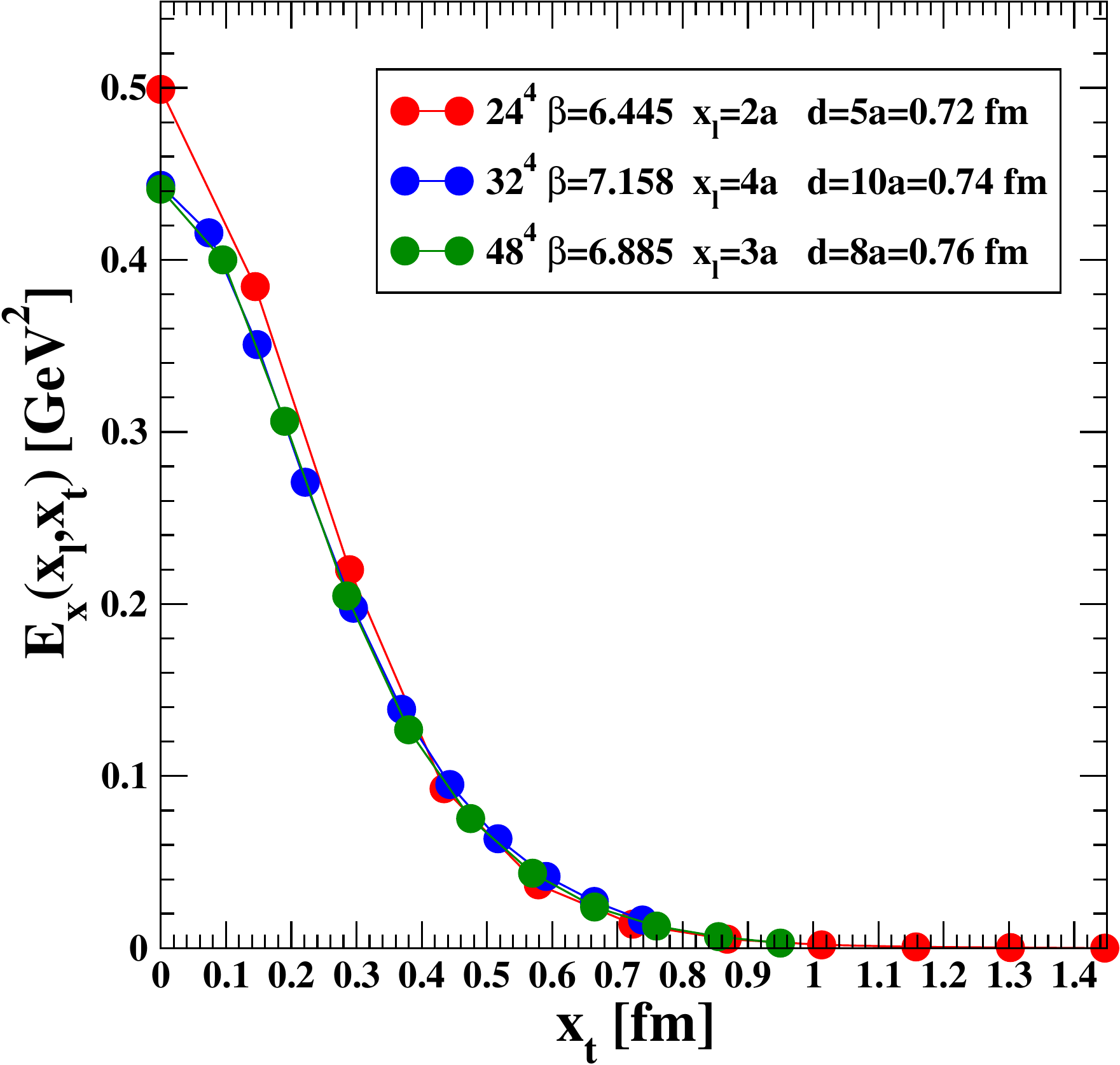}} \hspace{1 cm}
\subfigure[]{\includegraphics[width=0.4\textwidth,clip]{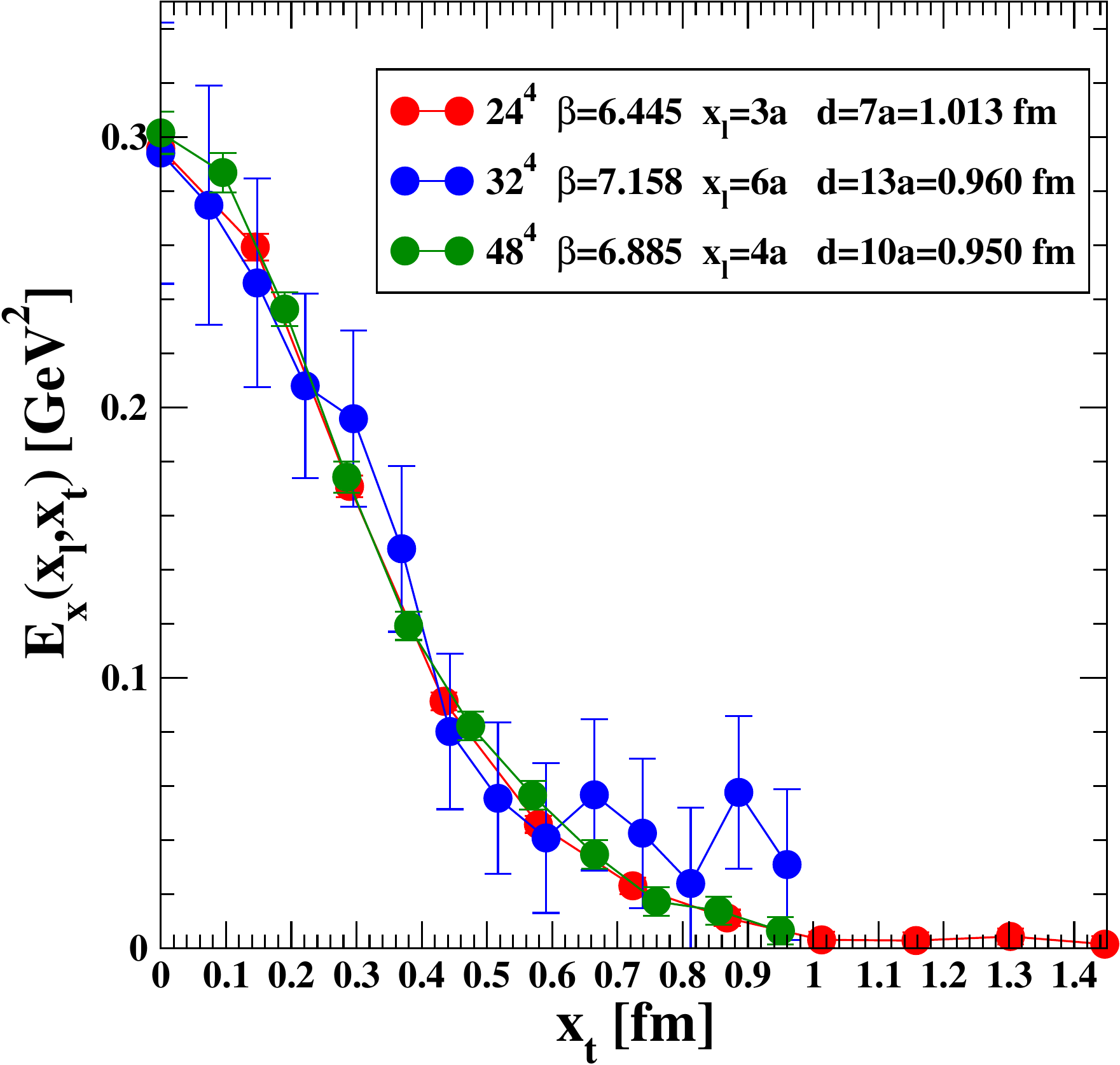}} 
\caption{
The value of the $E_x(x_l,x_t)$ field in physical units evaluated at midpoint of the line connecting the static quark-antiquark pair placed at the same distance physical distance $d$ for
several lattice volumes and several lattice distances: (a) $d\simeq0.75$~fm; (b) $d\simeq1$~fm. 
}
\label{fig:scaling}
\end{figure}

\begin{figure}[tb] 
\centering
\subfigure[]{\includegraphics[width=0.45\textwidth,clip]{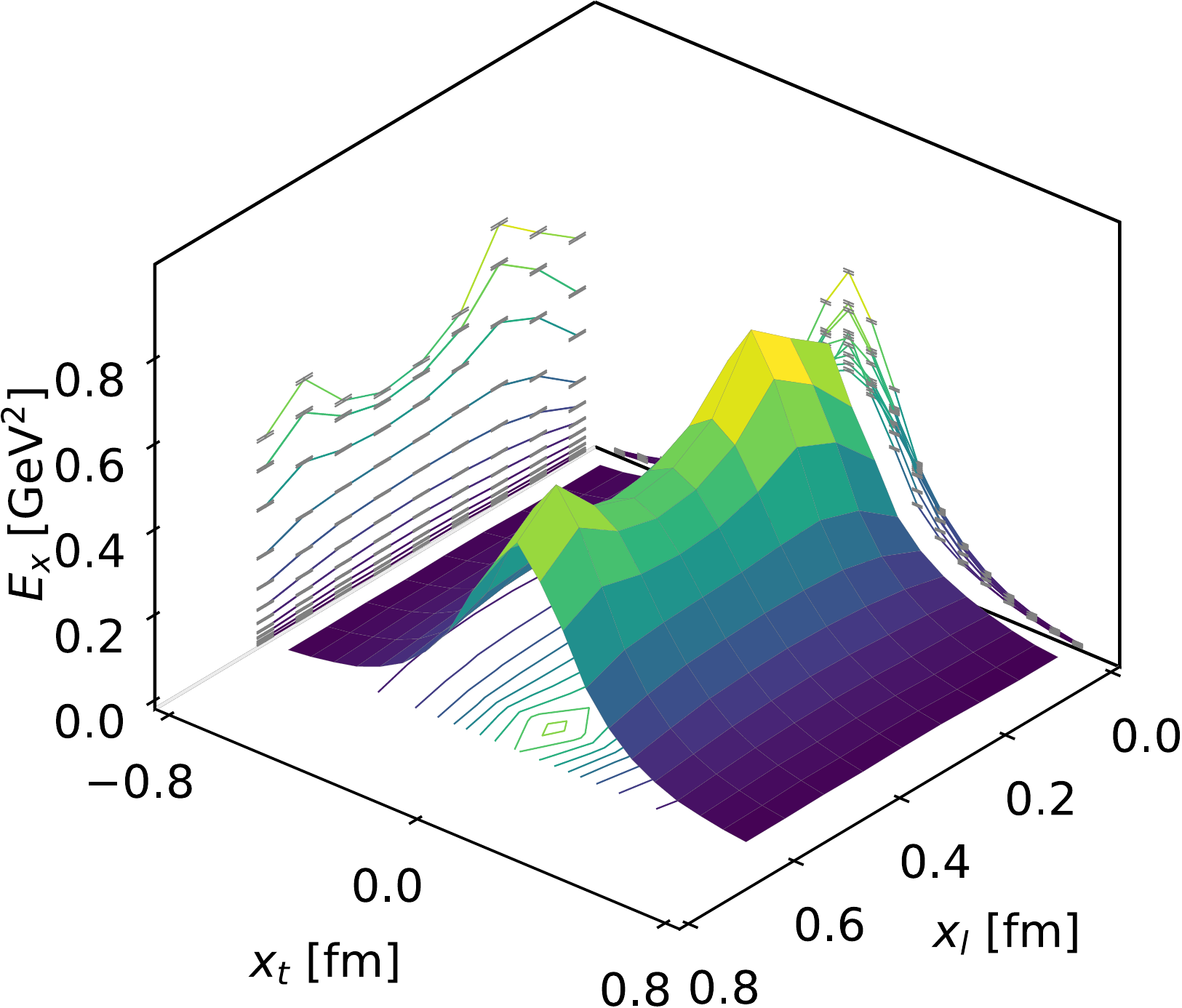}} \hspace{1 cm}
\subfigure[]{\includegraphics[width=0.45\textwidth,clip]{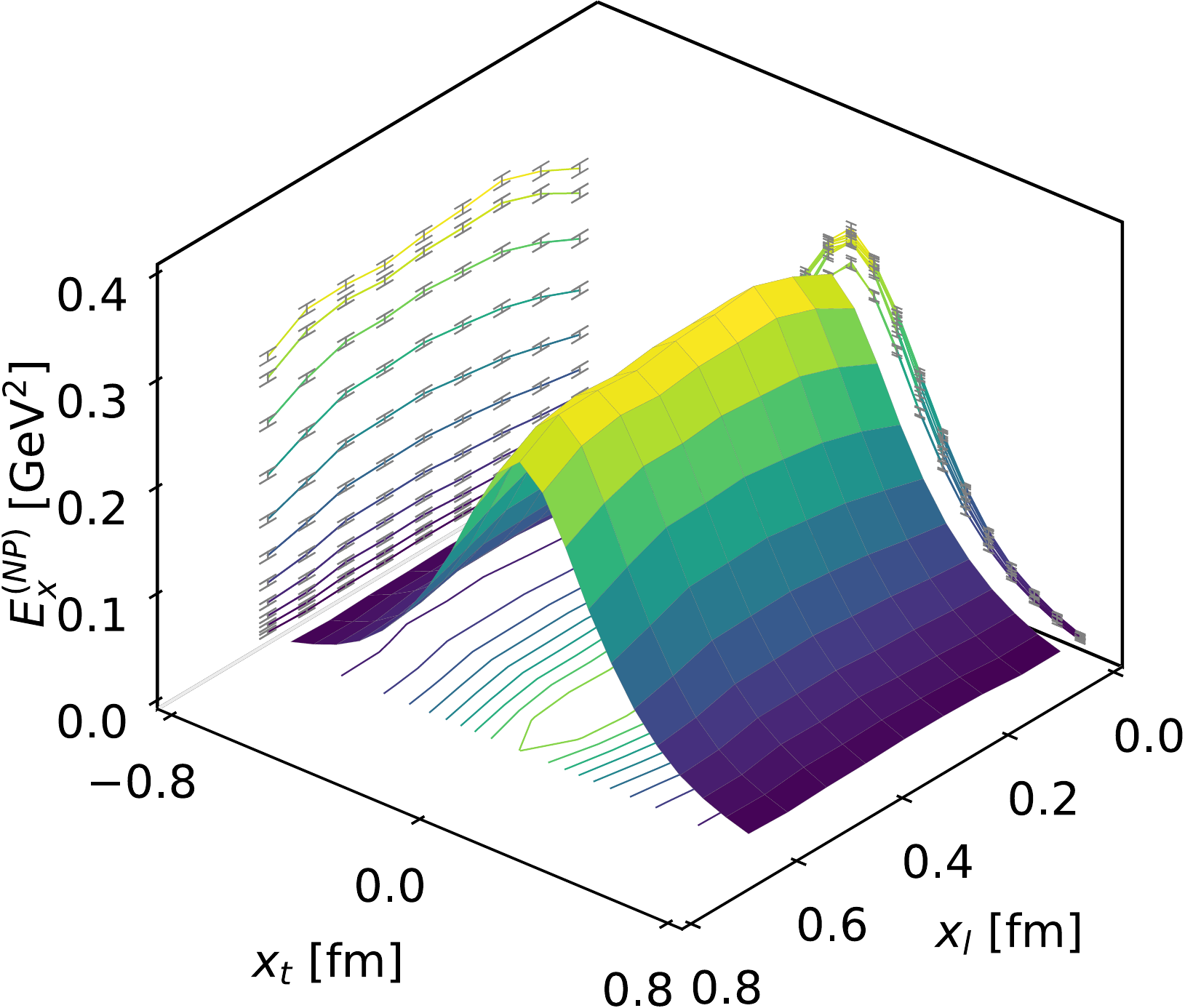}} 
\caption{
Surface and contour plots for the longitudinal components of the full (a)  and non-perturbative (b) chromoelectric field obtained by subtracting the perturbative contribution using
the procedure based on the irrotational condition Eq.~(\ref{eq:irrotational})  at $\beta=7.158$ and quark-antiquark pair at distance $d=10a=0.75$~fm.}
\label{fig:Exbeta7.158}
\end{figure}

We checked the physical scaling for several volumes and lattice spacings by measuring the fields in the case of physical distances $d\simeq0.75{\text{ fm}}$ and $d\simeq1{\text{ fm}}$ between quark-antiquark pair.
In Fig.~\ref{fig:scaling} the results of this scaling test show that the measured values of the $E_x$ field at different values of the coupling are well in agreement within statistical errors. The scaling test insures that in the range of parameters used our results satisfy continuum scaling.
As in the case of SU(3) pure gauge we find that: (i) the chromomagnetic field is everywhere much smaller than the longitudinal chromoelectric field and is compatible with zero within statistical errors;
(ii) the dominant component of the chromoelectric field is longitudinal; (iii) the transverse components of the chromoelectric field are also smaller than the longitudinal component, 
but can be matched to the transverse components of an effective Coulomb-like field $\vec{E}^C(\vec{r})$ satisfying the conditions:
\begin{enumerate}
\item The transverse component $E_y$ of the chromoelectric field is identified with the transverse component $E_y^C$ of the perturbative field
\begin{equation}
E^C_y \equiv E_y.
\label{eq:Ecoulomb}
\end{equation}
\item The perturbative field $\vec{E}^C$ is irrotational
\begin{equation}
\nabla \times \vec{E}^C = 0.
\label{eq:irrotational}
\end{equation}
\end{enumerate}
By imposing the condition of being irrotational, we are able to evaluate (for details see Ref.~\cite{Baker:2019gsi}) the perturbative contribution $E^C$ to the longitudinal chromoelectric field and to extract the non-perturbative confining longitudinal chromoelectric field:
 \begin{equation}
 E_x^{\rm{NP}}= E_x - E_x^C.
 \label{eq:monperturbative}
 \end{equation}
In Fig.~\ref{fig:Exbeta7.158} is shown the 3-dimensional profile of the chromoelectric field $E_x$ generated by a static quark-antiquark pair whose mutual distance is 0.74~fm.
At fixed value of $x_l$ along the line connecting the quark sources the value at transverse distance $x_t$ has been evaluated. The perturbative contribution to $E_x$ can be determined by means of the irrotational condition Eq.~(\ref{eq:irrotational}) and, after subtraction from the full chromoelectric field, we get the nonperturbative confining field $E_x^{\rm{NP}}$.
 
\begin{figure}[tb] 
\centering
\subfigure[]{\includegraphics[width=0.4\textwidth,clip]{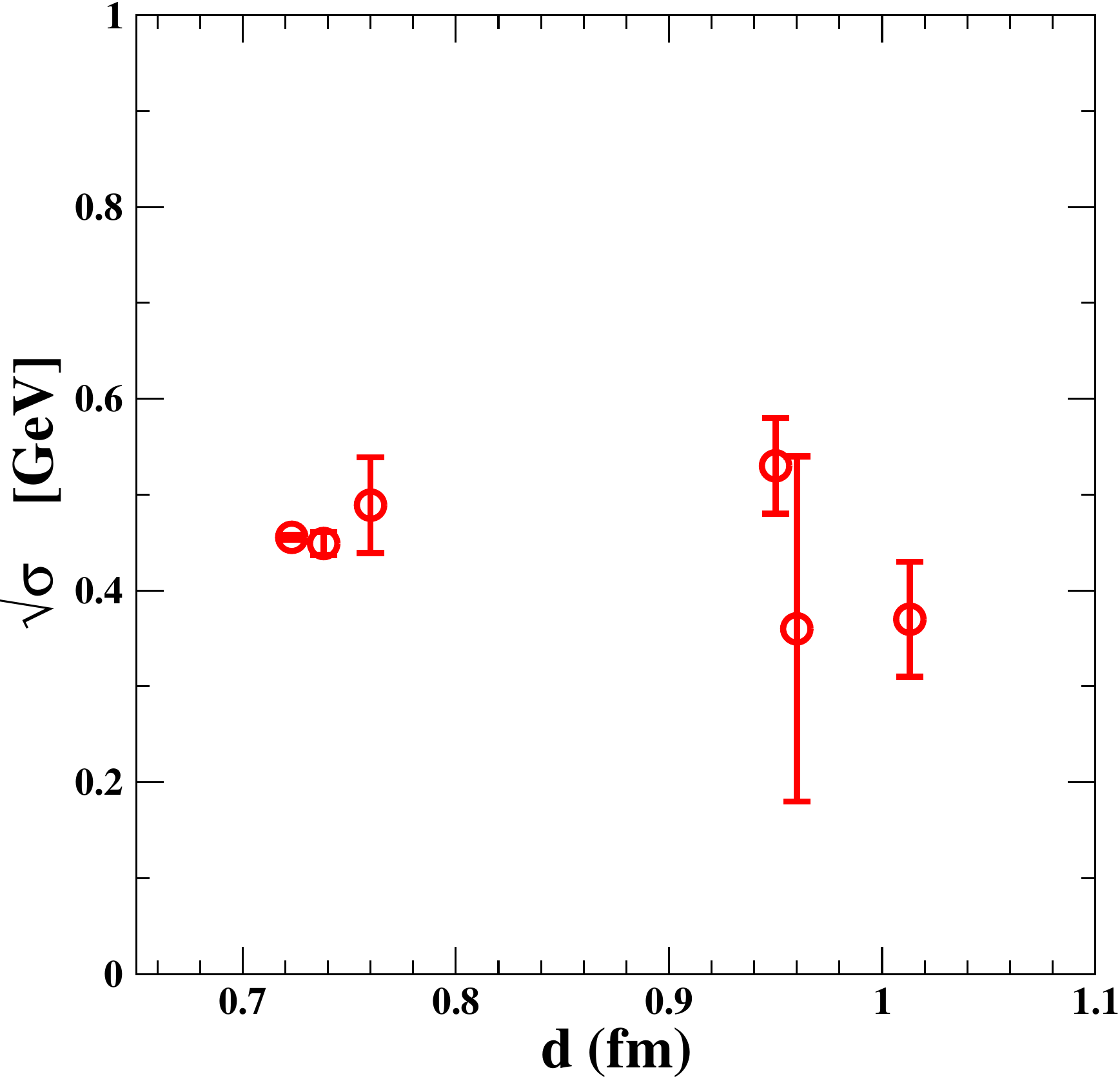}} \hspace{1 cm}
\subfigure[]{\includegraphics[width=0.4\textwidth,clip]{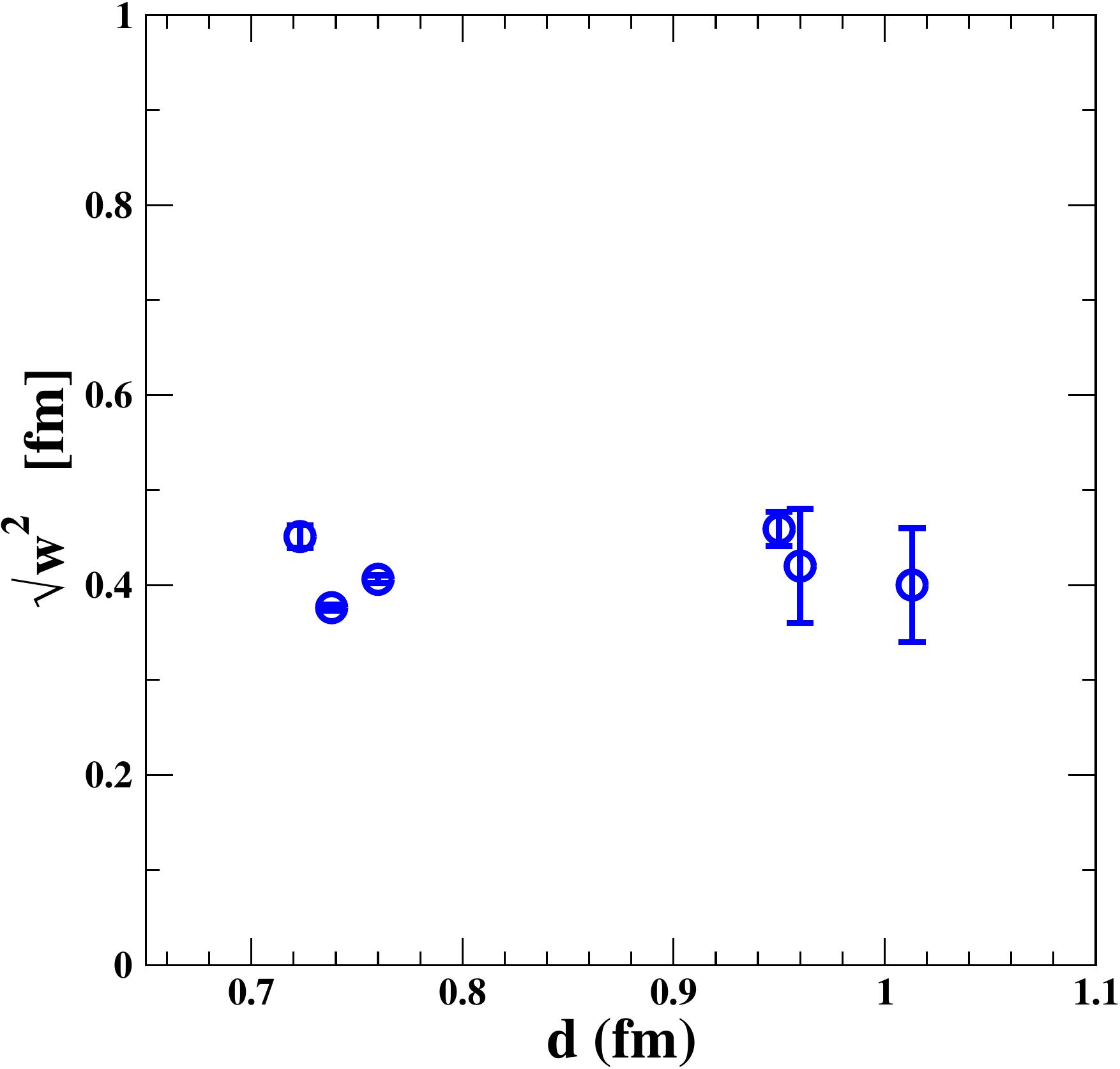}} 
\caption{
The square root of the string tension (a) and the root mean square width of the flux tube obtained in QCD with (2+1) HISQ flavors for several distances between the static quark-antiquark pair.}
\label{fig:stringwidth}
\end{figure}

 From the subtracted, nonperturbative part of the longitudinal chromoelectric field we can extract some relevant parameters of the flux tube at the midpoint, such as the root mean square width
 \begin{equation}
 \sqrt{w^2} = \sqrt{\frac{\int d^2x_t \, x_t^2 E_x(x_t)}{\int d^2x_t \, E_x(x_t)}},
 \label{eq:rmswidth}
 \end{equation}
  and the square root of the string tension
 \begin{equation}
 \sqrt{\sigma} =  \sqrt{ \int d^2x_t 
  \frac{(E^{\rm NP}_x)^2 (x_t)}{2} } .
 \label{eq:stringtension}
 \end{equation}
 We stress that the nonperturbative field $E_x^{\rm{NP}}$ was determined by a model-independent procedure and, therefore, all the results obtained from it are model independent as well.
 In Fig.~\ref{fig:stringwidth} we present our results for the  string tension and  the width  of the flux tube.
 This evaluation has been done for several values of the distance between the static quark-antiquark pair: results are quite independent from this distance. 
 A systematic study for several distances between the quark sources is in progress.

\section*{Acknowledgements}

This work was based in part on the MILC Collaboration's public lattice gauge theory code ({\url{http://physics.utah.edu/~detar/milc.html}).
Simulations have been performed using computing facilities at CINECA (INF20$\_$npqcd project under CINECA-INFN agreement).


\providecommand{\href}[2]{#2}\begingroup\raggedright\endgroup

\end{document}